\documentclass[prb,preprint,a4paper,floatfix]{revtex4-1}

\usepackage{latexsym}
\usepackage{graphicx}
\newcommand{\figwidtha}{6.0 in}
\newcommand{\figwidthb}{3.8 in}
\usepackage{verbatim}
\usepackage{amsmath}
\usepackage{color}

\begin{document}

\title{Structure and ferromagnetic instability of the oxygen-deficient SrTiO$_3$ surface}
\author{Soham S. Ghosh$^{(1)}$}
\author{ Efstratios Manousakis$^{(1,2)}$}
\affiliation{
$^{(1)}$ Department  of  Physics and National High Magnetic Field Laboratory,
  Florida  State  University,  Tallahassee,  FL  32306-4350,  USA\\
$^{(2)}$Department   of    Physics,   University    of   Athens,
  Panepistimioupolis, Zografos, 157 84 Athens, Greece
}
\date{\today}
\begin{abstract}
{\bf SrTiO$_3$ (STO) is the substrate of choice to grow oxide thin-films  
and oxide heterojunctions, which can form  quasi-two-dimensional electronic 
phases that exhibit a wealth of phenomena\cite{Thiel29092006,Ueno2008,Brinkman2007,Reyren31082007,Reyren31082007,Li2011,Bert2011}, 
and, thus, a workhorse in the emerging field 
of metal-oxide electronics\cite{Ramirez09032007,Cen20022009}. 
Hence, it is of great importance to know the exact 
character of the STO surface itself under various oxygen environments.
Using density functional theory within the spin generalized gradient 
approximation we have investigated the structural, electronic and 
magnetic properties of the oxygen-deficient STO surface. We find that
the surface oxygen vacancies order in periodic arrays giving rise to 
surface magnetic moments and a quasi 
two-dimensional electron gas in the occupied Ti 3-d orbitals. The surface confinement,  the oxygen-vacancy ordering, and the octahedra 
distortions give rise to spin-polarized \textit{t}$_{2g}$ dispersive sub-bands;  their energy 
split near the Brillouin zone center acts as an effective 
Zeeman term, which, when we turn on a Rashba interaction, produces bands with momentum-spin correlations similar to 
those recently discovered on oxygen deficient STO surface\cite{Santander-Syro2014}.}
\end{abstract}

\maketitle

The interface between two transition metal oxides forms
quasi-two dimensional  electronic 
phases\cite{Ohtomo427,Hwang2012,Cen2008,Mannhart26032010} 
that exhibit a rich set of phenomena, including tunable 
insulator-superconductor-metal transitions\cite{Thiel29092006,Ueno2008}, 
large magnetoresistance\cite{Brinkman2007}, 
ferromagnetism\cite{Brinkman2007} and superconductivity\cite{Reyren31082007,Li2011,Bert2011}.
The perovskite SrTiO$_3$ (STO)
is the substrate used for many such 
heterojunctions, and, therefore, it plays a crucial role
in the area of metal-oxide interfaces\cite{Ramirez09032007,Cen20022009}.
 Although it is a non-magnetic wide-band-insulator
 in bulk with a band-gap of 3.2 eV\cite{Ohtomo427},
 a quasi two-dimensional electron gas (Q2DEG) with interesting 
spin-momentum correlations has been detected 
 on its surface\cite{Santander-Syro2011,Plumb2014,Rice2014,Santander-Syro2014,Meevasana2011}. 
This Q2DEG can be produced through 
multiple methods, such as gating\cite{Ueno2008,Lee2011_1}, by creating surface
oxygen vacancies\cite{Santander-Syro2011,Plumb2014} or by
inserting a $\delta$-doped
 layer\cite{Kim2011,Kozuka2009,Kozuka2010,Jalan2010} inside the bulk crystal.

As seen by angle-resolved photoemission spectroscopy (ARPES), the Q2DEG has
 a strongly confined component made up of two sub-bands, 
 separated from each other by about 90 meV, forming concentric circular
 sections of the Fermi surface (FS)\cite{Santander-Syro2011,Meevasana2011}, and 
 a relatively less strongly confined component, about 200 meV higher in energy,
 forming ellipsoidal sections of the FS\cite{Plumb2014}.  
 Ab-initio\cite{Shen2012,Satpathy2008} and 
model\cite{Khalsa2012,Khalsa2013} calculations have identified 
Ti \textit{t}$_{2g}$ as the host states, though there are open 
questions about the nature of the Q2DEG. The two concentric 
rings of the FS
 on the \textit{k$_x$-k$_y$} plane are made up of  3\textit{d$_{xy}$} 
 orbitals, and two ellipsoids aligned along 
the \textit{k$_x$}  and \textit{k$_y$} 
 directions are 3\textit{d$_{xz}$/d$_{yz}$} states
 in nature\cite{Plumb2014}. The 3\textit{d$_{xz}$/d$_{yz}$} states are 
quasi one dimensional, and as seen from layer-by-layer ARPES 
study \cite{Chang2013} on LaTiO$_3$/SrTiO$_3$ interface, they 
have bulk-like behavior. These FS features have been seen to be created
by intense ultraviolet irradiation\cite{Meevasana2011}. They
persist under varying levels of oxygen vacancies 
and annealing conditions \cite{Plumb2014}, and, thus, seem ubiquitous. 

At Titanate interfaces with a polar perovskite, such as LaAlO$_3$, it has been 
suggested\cite{Pavlenko2013} that oxygen vacancies (Ov) create 
magnetic moments through orbital reconstruction and \textit{e}$_g$
occupancy, thereby modifying the occupied electronic levels produced 
by the so-called polar catastrophe. Oxygen-deficient
STO in bulk has been discovered to contain optically induced
magnetic moments\cite{Rice2014}, and the magnetism inverts sign
upon switching between right- and left-circular polarization.
Recently, a spin- and angle-resolved 
photoemission spectroscopy (SARPES) study\cite{Santander-Syro2014} has shown that
the concentric circular bands have a gap at the $\Gamma$ point, opposite spin chirality
and a large value of split at the Fermi surface that cannot be
explained by a purely Rashba-like model. It was suggested that 
this state breaks the time-reversal symmetry\cite{Santander-Syro2014}, and
an argument in favor of an antiferromagnetic ground state as 
the result of one of the instabilities of the 2D Fermi 
liquid with exchange interactions has been presented\cite{0953-8984-27-25-252001}.

In this paper, through a fully-relaxed spin-unconstrained  
DFT study of (001) terminated STO slabs with various surface oxygen
vacancy patterns, we show that stable periodic arrays of oxygen
 vacancies  can form in a surface TiO$_2$ layer
and cause the spin-degeneracy of the occupied sub-bands 
to be lifted at the GGA level giving rise to a weakly
ferromagnetic ground state. By considering slabs with 
multiple surface oxygen vacancy configurations (such as those of Fig.~\ref{fig:geometries} and those discussed in
Supplementary Note 1) and various surface Ov densities 
$n_{v}$ (= 1/4, 1/6, and 1/8),
we find that for a given surface Ov density, the superstructure in which the surface oxygen vacancies order in 
periodic stripe arrays, such as the one in Fig.~\ref{fig:geometries}, is 
characterized by the lowest energy. In addition,  Fig.~\ref{fig:O_def_stability} shows the particular 
striped ordered vacancy configuration  which is the most stable within 
an allowed range of oxygen chemical potential.
Furthermore, the Ti atoms 
 of the top-layer and the layer just below it acquire 
stable magnetic moments below surface Ov density of $n_v=1/8$
(refer to Supplementary Material Table~S1)
producing an internal Zeeman field, in which the level of doping and 
the surface magnetic moments
are functions of surface Ov density. 
At the $\Gamma$ point,
 the Q2DEG is spread over multiple layers and hosted 
by spin-non-degenerate states several hundred meV below the 
Fermi level. These deeper occupied states near the 
zone-center, which are strongly 3\textit{d$_{xy}$} in character,
host a Q2DEG but contribute 
a relatively small part to the net
magnetic moment, most of which is localized on the Ti atoms
at the oxygen-deficient surface away from the zone center,
and hosted by a reconstructed mixture of \textit{t}$_{2g}$ and
\textit{e}$_g$. Last, these findings via ab initio calculations 
are strongly supported by the following results. Taking into account a 
Rashba interaction
in the x-y plane, we find that these spin-split bands near the zone center
acquire the spin-momentum correlations
recently discovered\cite{Santander-Syro2014} by a SARPES study. 
The zone center gap 
is enhanced by the introduction of a Coulomb U term via a GGA+U calculation, 
but  the qualitative features of the spin-chirality
and the momentum-spin correlations of the bands in the direction parallel
to the surface, as well as the bands themselves
near the $\Gamma$ point are reproduced. The spin chirality perpendicular
to the surface is attributed to a 
matrix element of the SARPES process.\cite{Santander-Syro2014,Mirhosseini2013}

\begin{figure}[htp]
\vskip 0.2 in
\begin{center}
\includegraphics[width=\figwidthb]{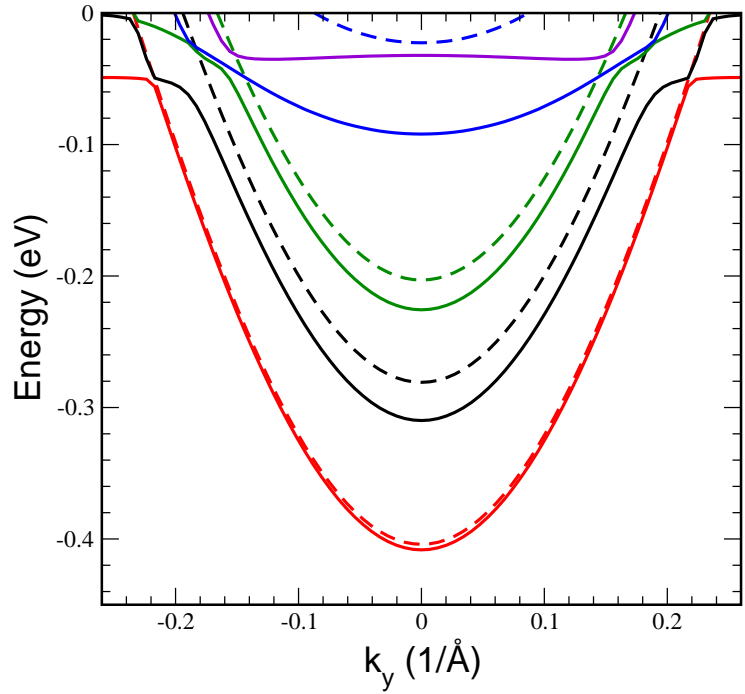}\\
\end{center}
\caption{(color-online) The near-$\Gamma$ 
  spin-GGA band structure of the
   STO slabs with striped vacancy configuration $(n_{v} = 1/6)$.
   Up-spins bands are denoted with solid lines 
  and corresponding down-spin bands denoted with dashed lines. 
  The 3\textit{d}$_{xy}$ bands are light and dispersive, while the  3\textit{d}$_{xz}$
  and the 3\textit{d}$_{yz}$ are heavier and less dispersive due
  to confinement in the z-direction, which also produces the observed level
  quantization of states. 
}
\label{fig:gammabands}
\vskip 0.2 in
\end{figure}

The Q2DEG
in STO has been associated with a band bending of 
$\sim 300$ meV near the 
surface\cite{Santander-Syro2011,Meevasana2011}.
Confinement in this potential-well lifts the degeneracy 
between the 3\textit{d}$_{xy}$ of different Ti atoms and 
produces a series of light parabolic sub-bands split in energy, 
which produce concentric rings on the Fermi surface
near the zone center, once doped. Oxygen vacancy doping 
lifts the spin-degeneracy in each of these bands but
by different magnitudes, as shown in Fig.~\ref{fig:gammabands}. 
We find that four STO-layer and six STO-layer slabs have qualitatively similar
band structure 
(this is illustrated in Fig.~S2 of Supplementary Material),
hence we believe that our results and conclusions
presented here, which are obtained from slabs containing only four STO-layers,
may be applicable to the semi-infinite system.
At 1/6 surface Ov density, 
these states are as far deep as $400$ meV below the Fermi 
level at the $\Gamma$ point. 
Their site and $Y_{lm}$-projection 
(refer to Supplementary Material Note 3) 
illustrate that 
each of the spin-split pair of the light sub-bands has 
a preference for a Ti orbital belonging to a different TiO$_2$ layer.
This is an effect of the confining potential-well that quantizes levels
in the (001) direction. 
Around the $\Gamma$-point, they are all 3\textit{d}$_{xy}$, 
except the heavy sub-bands $\sim 50$ meV below the Fermi surface, 
which are largely 3\textit{d}$_{xz}$.
At the GGA level, there are multiple spin pair bands 
(see Fig.~\ref{fig:gammabands} and  Supplementary Material Fig.~S3) 
 that have significant contribution from the Ti atoms
immediately  below the oxygen deficient surface,
and within the range of a SARPES experiment. However,
as we illustrate in the Supplementary Material (Fig.~S4), 
addition of an on-site Hubbard Coulomb-U term causes the electrons to 
repel each other and move away from the surface, leaving 
all but one pair within the range of the SARPES probe.
This pair of bands associated with atoms near the TiO$_2$ surface  
has a zone-center splitting of $\sim 29$ meV at the GGA
level and $\sim 80$ meV at the GGA+U (U = 2 eV) level.


 By switching on the spin-orbit coupling (SOC) in the system, 
 we find a slightly modified near-$\Gamma$ band structure 
for the four  STO-layer slab, as shown in the
Supplementary Material Fig.~S5.
  The change in the 3\textit{d}$_{xy}$
  bands around the $\Gamma$ point is negligible. 
  The biggest effect of SOC can be seen in lifting the degeneracy of the
band-crossing at the corners, as expected. 
 It preserves the net magnetic 
 moment at the surface and the spin-polarization of the 
 3\textit{d}$_{xy}$ bands around the $\Gamma$ point. At this
mean-field level, SOC effect is predictably
 small enough so that the spin-orbit mixing changes the spin-eigenstates
only by a small amount and the spin up-down characterization remains valid.
The most important role of the spin-orbit coupling
is that it simply selects the direction of the effective internal ``Zeeman
field'' (which splits our lowest energy bands within our GGA-spin 
calculation)
to be perpendicular to the orbital angular momentum $x-y$ plane.

\begin{figure}[htp]
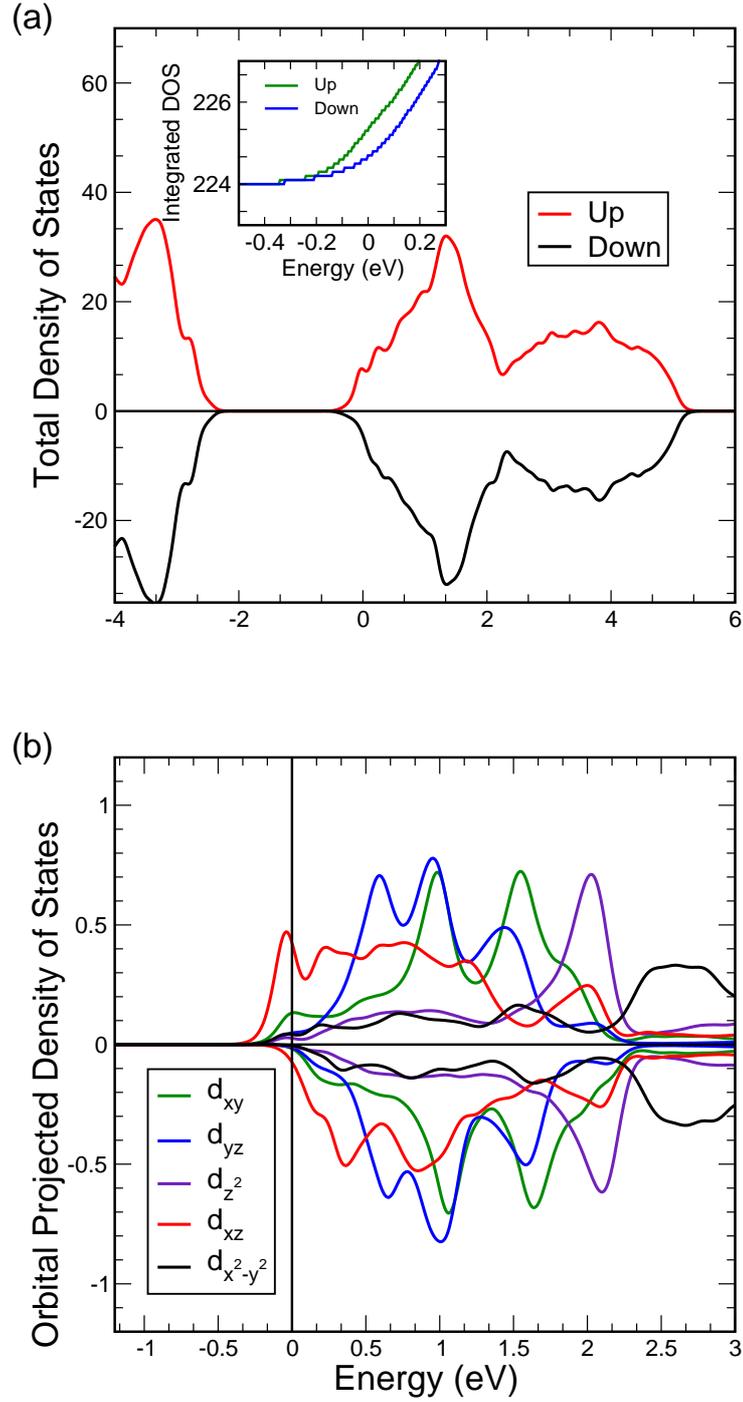

     \begin{center}
\includegraphics[width=\figwidthb]{Fig4a.eps}\\
\vskip 0.3 in
\includegraphics[width=\figwidthb]{Fig4b.eps}
\end{center}
\caption{(color-online) (a)
  Total up and down DOS of a four STO-layer slab with
  1/6 Ov surface density at the TiO$_2$ surface.
  The zero is the Fermi level at the
  present level of doping. 
  (inset) Integrated up and down dos, showing the onset of magnetism
  at $\sim 300$ meV below the Fermi level.
(b)
 Projected DOS of the 3d orbitals of a surface Ti atom next to an oxygen vacancy.
  The occupied polarized states are a mixture of spin-polarized \textit{t}$_{2g}$
  and \textit{e}$_g$ located around the S-point of the BZ.}
\label{fig:dos}
\vskip 0.2 in
\end{figure}

Fig.~\ref{fig:dos} shows the total density of states (DOS) of a 
four STO-layer slab along with the 
integrated up and down DOS as a function of energy
as well as the orbital projected DOS of one of 
the Ti atoms at the surface 
at the GGA level.
We find that in the absence of a Rashba term
and at 1/6 Ov surface density,
the low-lying 3\textit{d}$_{xy}$ bands near the
zone-center carry finite but small amounts of magnetic moments. 
Most of the contribution to the magnetic moment comes from heavier
sub-bands away from the zone-center, and depend on a reconstruction of the
\textit{e}$_g$ states. 
As we can observe in Fig.~\ref{fig:dos}, the spin-polarized \textit{e}$_g$
states of the Ti atom next to a missing oxygen is
pulled below the Fermi level and get 
mixed with the \textit{t}$_{2g}$ states. 
On the other hand, the occupied bands of the Ti atoms one 
layer below the surface
are  entirely \textit{t}$_{2g}$ in nature with almost 
non-magnetic 3\textit{d}$_{xy}$
states (refer to Supplementary Material Fig.~S6).

The splitting of the \textit{e}$_g$ states and their
mixing with the \textit{t}$_{2g}$ states give the surface Ti atoms
a magnetic moment  (with magnetic moments for the two nonequivalent surface Ti atoms given
by $M_{Ti^{(1)}} = 0.118 \mu$B and $M_{Ti^{(2)}} = 0.228 \mu$B) each for $n_v=1/6$.

\begin{figure}[htp]
\vskip 0.2 in
     \begin{center}
            \includegraphics[width=\figwidthb]{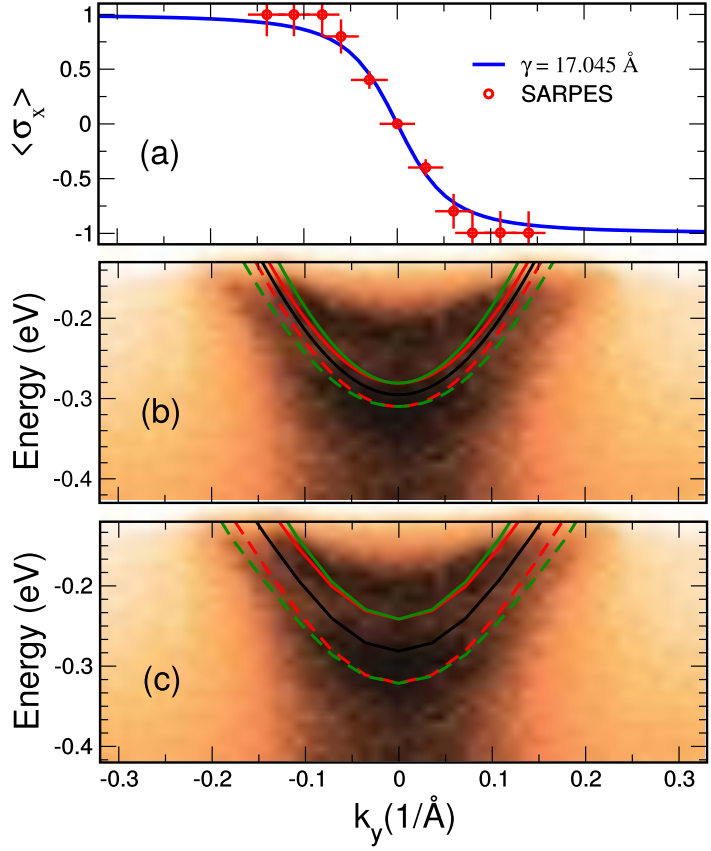}
\end{center}
\caption{(color-online) 
(a) The result of fitting 
the experimentally determined\cite{Santander-Syro2014}
spin x-component to the form given by Eq.~\ref{sigma}
indicating a large planar Rashba coupling near the 
surface. 
(b) The near-surface GGA bands of the four STO-layer slab after applying 
a Rashba spin-split  term using Eq.~\ref{energy} 
as compared to the experimental
results\cite{Santander-Syro2014}. 
(c) The same as in (b) using the bands by GGA+U.
The black solid line is the result of GGA or GGA+U without spin-relaxation.
The green and red solid (dashed) lines are obtained with GGA or GGA+U 
after spin-relaxation for down (up) with and without the Rashba interaction 
respectively. } 
\label{fig:rashba_fit}
\vskip 0.2 in
\end{figure}

The band picture shown in Fig.~\ref{fig:gammabands} 
 does not take into  consideration the Rashba effect, which couples 
spin with momentum. In the existence of strong 
fields caused by surface termination and oxygen 
vacancy, inversion symmetry is lost and the octahedra buckle,
causing previously forbidden hopping channels 
to open. This, coupled with a strong spin-orbit 
coupling\cite{Khalsa2013} produces a prominent planar Rashba effect.
For the Ti atom one unit-cell
below the oxygen deficient surface, this effect can be
modeled using a simple 2-D Hamiltonian for small values
of \textbf{k}. Written in the basis of Ti 3\textit{d}${_{xy}}$ 
Bloch states and $S_{z}$ eigenstates, the Hamiltonian is:
\begin{equation}
  \hat{H} =   \epsilon_{GGA}(\vec k) \hat 1  -
\alpha(\sigma_{x}k_{y} - \sigma_{y}k_{x}) - h \sigma_{z},
\label{eq:Hamiltonian}
\end{equation}
where $\alpha$ is the Rashba parameter, $\sigma_{x}$, $\sigma_{y}$
and $\sigma_{z}$ are Pauli matrices, $\hat 1$ is the unit matrix, 
and the last term is the Zeeman term due to an effective internal mean field
in the z-direction which is responsible for the spin-splitting the 
bands which we found by allowing for spin- and lattice relaxation 
within the GGA and GGA+U. In the absence of an external magnetic field, 
both magnetic directions along the z-axis are equally likely, and we
use $h = \pm|h|$ to account for ferromagnetic domains that average to 
zero in a macroscopic measurement like SARPES\cite{Santander-Syro2014}.
 Here $\epsilon_{GGA}(\vec k)$ is the band structure obtained 
as the average between the lowest two near-surface bands closer to the oxygen deficient TiO$_2$ surface obtained in 
our DFT  calculation  which correspond
to spin-up and spin-down states, and
the value of this field will be chosen to 
reproduce the energy splitting of these two lowest bands.
In addition, $\epsilon_{GGA}(\vec k)$ is very close (apart from a constant)
to the results of our DFT calculation when we do not include spin.
The eigenvalues of this Hamiltonian are
\begin{equation}
e_{\pm}(k) = \epsilon_{GGA}(\vec {k}) \pm |h|\Delta(k);\hskip 0.2 in 
\Delta(k) = \sqrt{1 + \gamma^{2}k^{2}}. 
\label{energy}
\end{equation}
where $k^2 = k_x^2 + k_y^2$ and $\gamma  = \alpha/|h|$.
The corresponding expectation values of the spin components are as follows:
\begin{eqnarray}
\left(\begin{array}{c}
\langle \pm | \sigma_{x}| \pm \rangle \\
\langle \pm | \sigma_{y}| \pm \rangle \\
\langle\pm | \sigma_{z} | \pm \rangle 
\end{array}\right)
 = 
\mp\frac{1}{\Delta(k)} \left(\begin{array}{c}
\gamma k_y \\
-\gamma k_x \\
{{h} \over {|h|}}
\end{array}\right).
\label{sigma}
\end{eqnarray}
In Fig.~\ref{fig:rashba_fit} we compare the experimentally 
determined  bands to those obtained using Eq.~\ref{energy} by 
choosing the  parameters in the following way.
Fig.~\ref{fig:rashba_fit}(a) shows the result of our fit 
of the experimentally determined 
x-component of the spin to the form given by Eq.~\ref{sigma} where the only
fitting parameter  is $\gamma$. The solid line is the result of
our fit with $\gamma =17.045$ \AA.
Fig.~\ref{fig:rashba_fit}(b) is obtained
using $h=14.55$ $\mathrm{meV}$ as
found by our spin GGA calculation and the value of $\gamma = 17.045$ \AA\,
determined by the previously discussed fit. This implies that
the Rashba parameter is $248$ $\mathrm{meV}$\AA\, for the given value of $h$.

Fig.~\ref{fig:rashba_fit}(c)  is obtained 
using $h=40.0$ meV as found in GGA+U calculation 
for a four STO-layer slab 
and $\alpha = 300$ $\mathrm{meV}$ \AA,\, illustrating a better agreement
of the zone-center gap and the 
momentum dispersion of the bands with experiment.

We note that
there is a possibility that SARPES overestimates 
the spin polarization by not accurately estimating the quasiparticle 
spectral weight away from the peak of the 
spectral function: it is clear that the quasiparticle 
function is widely spread around the peak and, thus, the
part of the spectral weight away from the peak can hide under the 
contribution of tails from other nearby bands. In such case 
the estimation of the Rashba parameter can be significantly 
affected. 

The  photo-excitation process itself mixes states with different angular
momentum components, which depend on the light polarization, causing 
a rotation of
the electron spin-polarization through the  spin-orbit coupling term. 
Thus, the expectation value $\langle \sigma_z \rangle $ picks up a contribution 
proportional to $\langle \sigma_x \rangle$ and/or to $\langle \sigma_y \rangle$.
In the absence of an external magnetic field, the average  value 
of the term proportional to $h$ over the entire surface
is expected to be zero. 
 This can explain that the expectation value of 
$\langle \sigma_z \rangle $ measured in the SARPES study has a 
$k$-dependence similar to
that of $\langle \sigma_x \rangle $ or $\langle \sigma_y \rangle $.

In conclusion, we find that our ab initio calculations
predict very interesting patterns of Ov ordering on the (001) STO surface which 
lead to a small ferromagnetic moment of the surface Ti atoms.
Furthermore,  major experimental 
characteristics of the STO 
surface can be qualitatively reproduced by a GGA (and GGA+U) spin  
calculation where we have included the
effects of the planar Rashba coupling. 
The underlying symmetry breaking and the presence of a quantum-well  
potential near the surface give rise to discrete energy bands in the (001) 
direction. When surface  oxygen vacancies are introduced and the structure is 
relaxed within a spin-dependent GGA, bands near the TiO$_2$ surface emerge
characterized by a dispersion near the zone center similar to that
observed by recent SARPES studies. 
More importantly, we find that the fully relaxed GGA calculation 
opens a spin-gap at the zone center, which is 
further enhanced by adding a moderate  U in a GGA+U calculation.
The contribution to the magnetic moments come primarily from 
the neighborhood of the $S$ and $X$ points 
in the BZ, and localized at the reconstructed $t_{2g}$ and $e_g$ of 
the Ti atoms at the oxygen deficient surface (Supplementary Material Fig.~S7).
At the spin-GGA
level and at the present high levels of doping, the system is 
weakly ferromagnetic.  
Therefore, it is reasonable to assume a theoretical framework whereby we identify
with these $3d_{xy}$ bands those seen in the most recent SARPES study\cite{Santander-Syro2014}. Then 
by taking these bands as  a starting point we 
turned on a planar Rashba term,
the lowest lying, closest to the TiO$_2$ surface 3\textit{d}$_{xy}$ bands
 qualitatively reproduce the 
experimental SARPES results\cite{Santander-Syro2014} for the energy bands
close to the $\Gamma$ point. Furthermore, they reproduce
the momentum-spin correlations in the $x$ and $y$ directions seen in SARPES and they
have no net magnetic moment in the $x-y$ plane.  

\section{Methods}
\label{method}
We consider slabs between four and six STO-layer thick, 
terminated in the (001) direction by 
an SrO surface on one end and an oxygen-deficient 
TiO$_2$ surface on the other. Such a
 system lacks a plane of inversion around the $x-y$ plane. 
The presence of surfaces in the (001) direction 
of our slab geometry creates a electrostatic potential which has 
one minima near the oxygen deficient
TiO$_2$ surface, and another near the TiO$_2$ layer
 on the opposite end. We focus on the 
dispersive bands localized close to the oxygen deficient surface
and within the depth-range of the ARPES study.

All computations were performed using the plane-wave basis set 
(plane wave cutoff of 540 eV) with the projector augmented wave (PAW) 
methodology\cite{Blochl} used to describe the wavefunctions of the core 
electrons, as implemented in the \textsc{VASP} package\cite{Shishkin3,Fuchs,Shishkin2,Shishkin1}. 
The Perdew-Burke-Ernzerhof (PBE) exchange correlation 
functional\cite{PBE} was used for all GGA calculations. 
The 4\textit{s} and 3\textit{d} electrons of the transition 
metal atom and the oxygen 2\textit{s} and 2\textit{p} electrons 
were treated as valence electrons.
An initial unit cell length of 3.944 \AA\ was used, which was found by 
curve-fitting the ground state energies of bulk STO to find the minimum.  
The Brillouin zone was sampled with a maximum of 
$5\times 15\times 1$ k-point mesh for the self-consistent cycles\cite{MP76}. Forces 
were converged to less than 10 meV/\AA\, for each ion. 
A local Coulomb repulsion of Ti 3$d$ electrons was accounted within 
GGA+U approach with $U_\textrm{Ti} = 2$ eV.

\vskip 0.2 in
\section{Acknowledgments}
The authors would like to thank C. S. Hellberg for useful communication.
This work was supported in part by the U.S. National High Magnetic Field
Laboratory, which is partially funded by the NSF DMR-1157490 and 
the State of Florida.

\newpage

\begin{figure}[htp]
\includegraphics[width=\figwidtha]{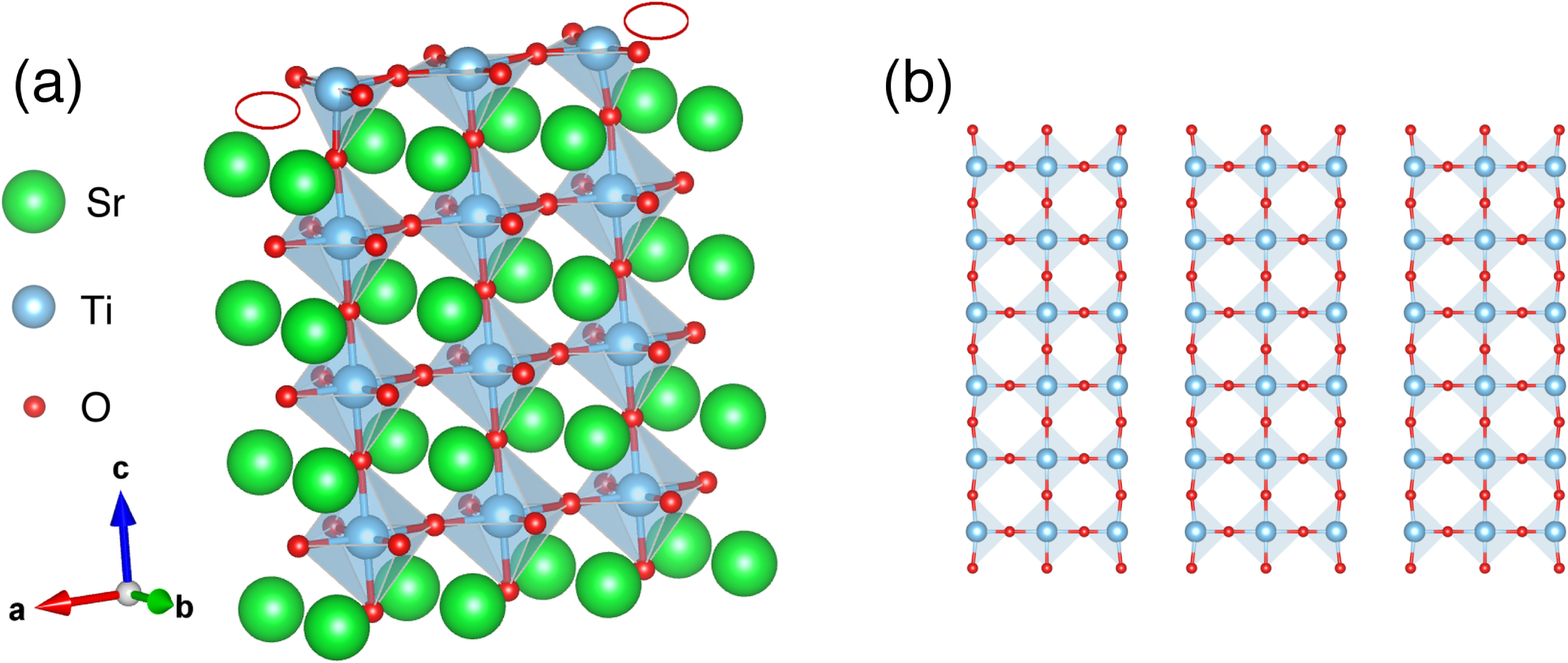}
\vskip 0.2 in
\caption{(color-online) (a) Four  
        layers  of atomically relaxed STO  
        bounded in the (001) direction by an SrO 
        surface on one side and a TiO$_2$ surface on the 
        other side with an imposed 1/6 surface Ov coverage. 
        The presence of the oxygen vacancies (the sites of which are denoted
by the two red circles) cause the tetrahedra at the TiO$_2$ surface to buckle
        and push both the O and the Ti atoms away from the
        lower layers. The Ti-Ti bond length at the surface is larger by about
        0.3 \AA \, along the x-axis as compared to that along the
        y-axis. 
        (b) Top surface head-on view of the superlattice structure of oxygen 
vacancy ordering in periodic arrays.
}
\label{fig:geometries}
\vskip 0.2 in
\end{figure}

\begin{figure}[htp]
\vskip 0.2 in
\begin{center}
\includegraphics[width=\figwidtha]{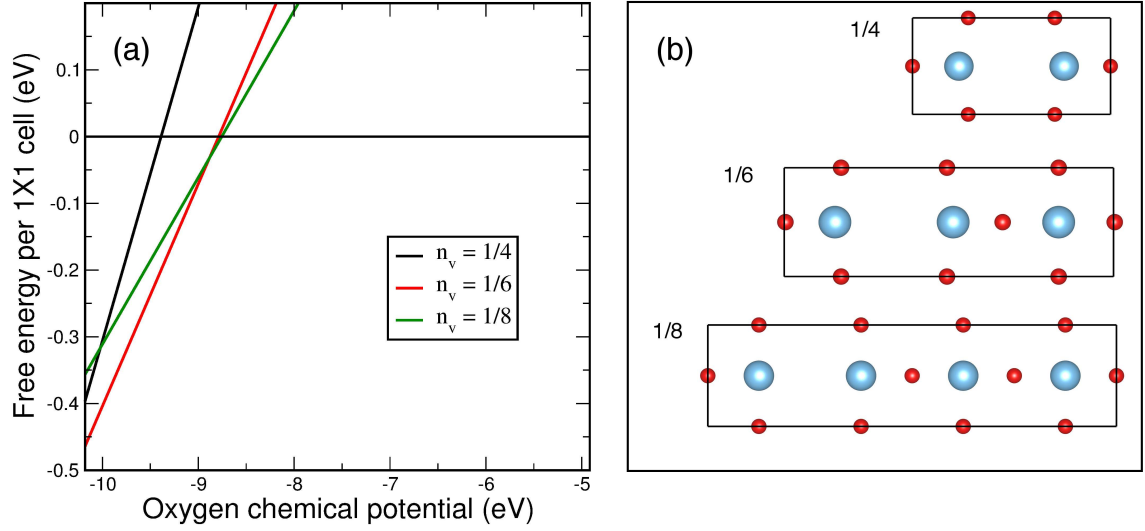}
\end{center}
\caption{(color-online)(a) Stability graph of the striped vacancy pattern
         discussed in Fig.\ref{fig:geometries} with different surface Ov 
         densities. All three densities produce similar sub-surface 
         3\textit{d$_{xy}$} bands near the zone-center but differ in the
         magnitude of the magnetic moments. For an explanation of the free energy
         and the allowed range of oxygen chemical potential see 
        Supplementary Material Note 6. 
         (b) Top view of the various Ti$_{(m)}$O$_{(n)}$ units at the TiO$_2$ surface
        with one oxygen vacancy per u.c.,
         showing  (i) a $2\times1$ u.c. with one eliminated O atom at (0.50, 0.50) 
         creating a stripe of missing oxygen atoms along the y-axis (Ov surface density $n_{v} = 1/4$)
          and giving each Ti atom
         exactly one nearest neighbor vacancy, (ii) a $3\times1$ u.c. with
         one eliminated O at (0.33, 0.50) causing a similar vacancy stripe
         but with lower density $(n_{v} = 1/6)$,
         and  (iii) a $4\times1$ u.c. with one eliminated O at (0.25, 0.50), creating
         vacancy stripes with a Ov surface density of $n_{v} = 1/8$.
         Further vacancy configurations are discussed in the 
         Supplementary Material.}
\label{fig:O_def_stability}
\vskip 0.2 in
\end{figure}

\vskip 0.2 in

\end{document}



\title{Supplementary Material\\[1cm]for the paper entitled\\[1cm]
Structure and ferromagnetic instability of oxygen-deficient SrTiO$_3$ surface}

\author{Soham S. Ghosh$^{1}$}
\author{ Efstratios Manousakis$^{1,2}$ }
\affiliation{
$^{(1)}$ Department  of  Physics and National High Magnetic Field Laboratory,
  Florida  State  University,  Tallahassee,  FL  32306-4350,  USA\\
$^{(2)}$Department   of    Physics,   University    of   Athens,
  Panepistimioupolis, Zografos, 157 84 Athens, Greece
}
\date{\today}
\maketitle
\beginsupplement
\section{Supplementary Note 1: Other vacancy configurations}
\begin{figure}[htp]
\vskip 0.3 in
\begin{center}
\includegraphics[width=\figwidthw]{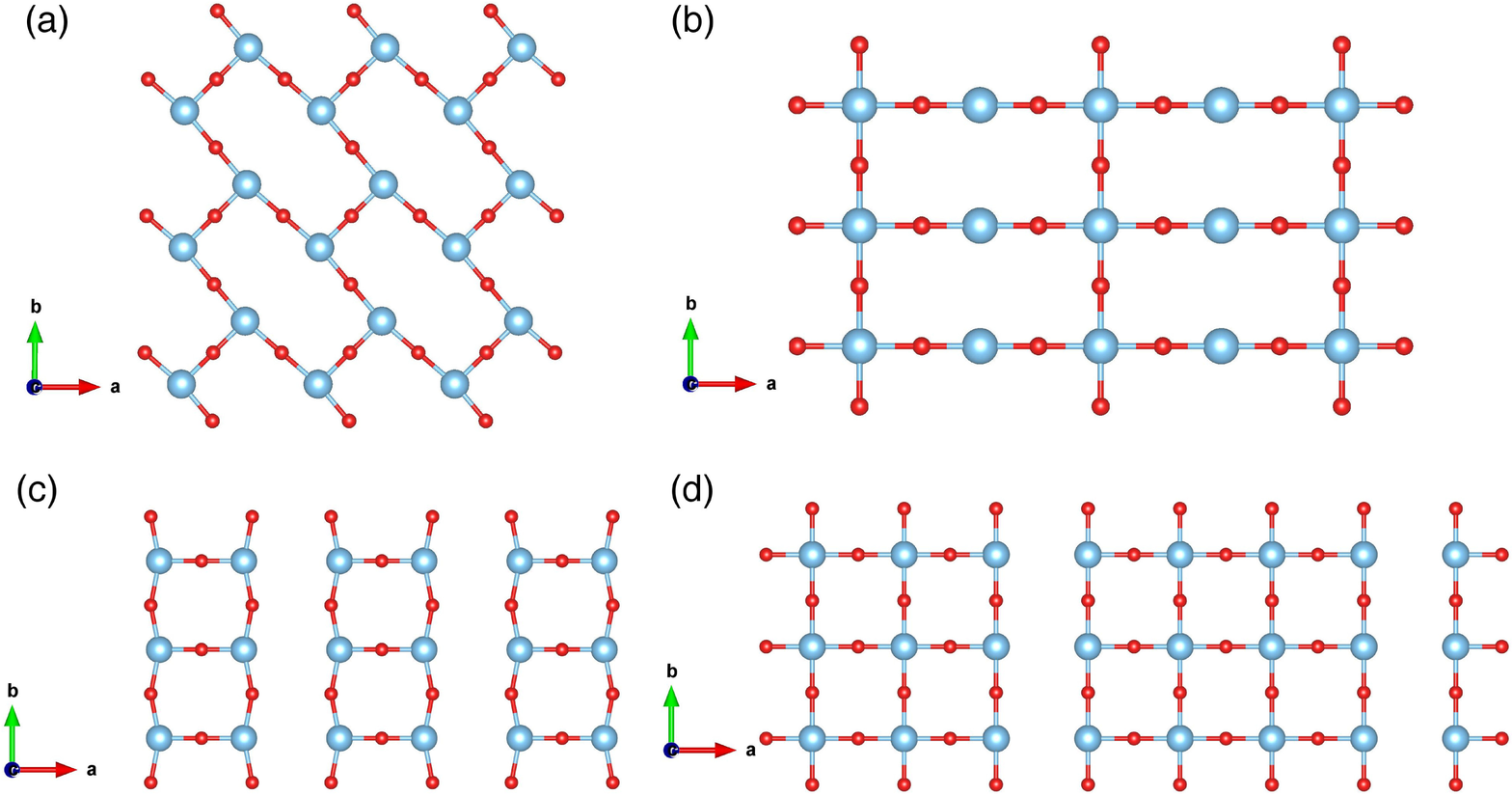}
\end{center}
\caption{(color-online) 
  (a) A  four STO-layer slab with  $\sqrt{2}\times\sqrt{2}$ (dimerized) 
         vacancies, causing each surface Ti to have one nearest neighbor vacancy.
          (b) A $2\times1$ configuration with 
        chains of vacancies in the y-direction. The configuration (b) 
contains two different types of surface Ti,
       one with no nearest neighbor vacancies, and the other with two nearest neighbor vacancies. 
       Both configurations have doping level $n_{v} = 1/4$ ($25\%$).
      (c) Stripe configuration, $n_{v} = 1/4$ and (d) 
stripe configuration, $n_{v} = 1/8$. 
}
\label{fig:S_unit_cells}
\end{figure}

Fig.~\ref{fig:S_unit_cells} shows the vacancy configurations
considered here which are not shown in the main part of the paper associated with Fig.~1. 
The non-stripe configurations are higher in energy 
by $\sim 0.5$ eV (for $n_{v} = 1/4$) per supercell used in the DFT calculations than the stripe configuration. 
The (1/4) stripe and dimer configurations have stable magnetic moments at the spin-GGA level 
while the (1/8) stripe and $2\times1$ configurations have no noticeable magnetic moments.
In Table~\ref{table:moments} we list the  magnetic moments of the 
surface and sub-surface Ti atoms for the systems under consideration. 
For the stripe configuration with $n_{v}$ = 1/4, the two surface Ti atoms have similar environment
and equal magnetic moments. For $n_{v}$ = 1/6, the Ti atoms next to the Ov has
less magnetic moment than the Ti with all nearest neighbor oxygen atoms present.

\section{Supplementary Note 2: The semi-infinite approximation}
\begin{figure}[htp]
\vskip 0.1 in
\begin{center}
\includegraphics[width=\figwidth]{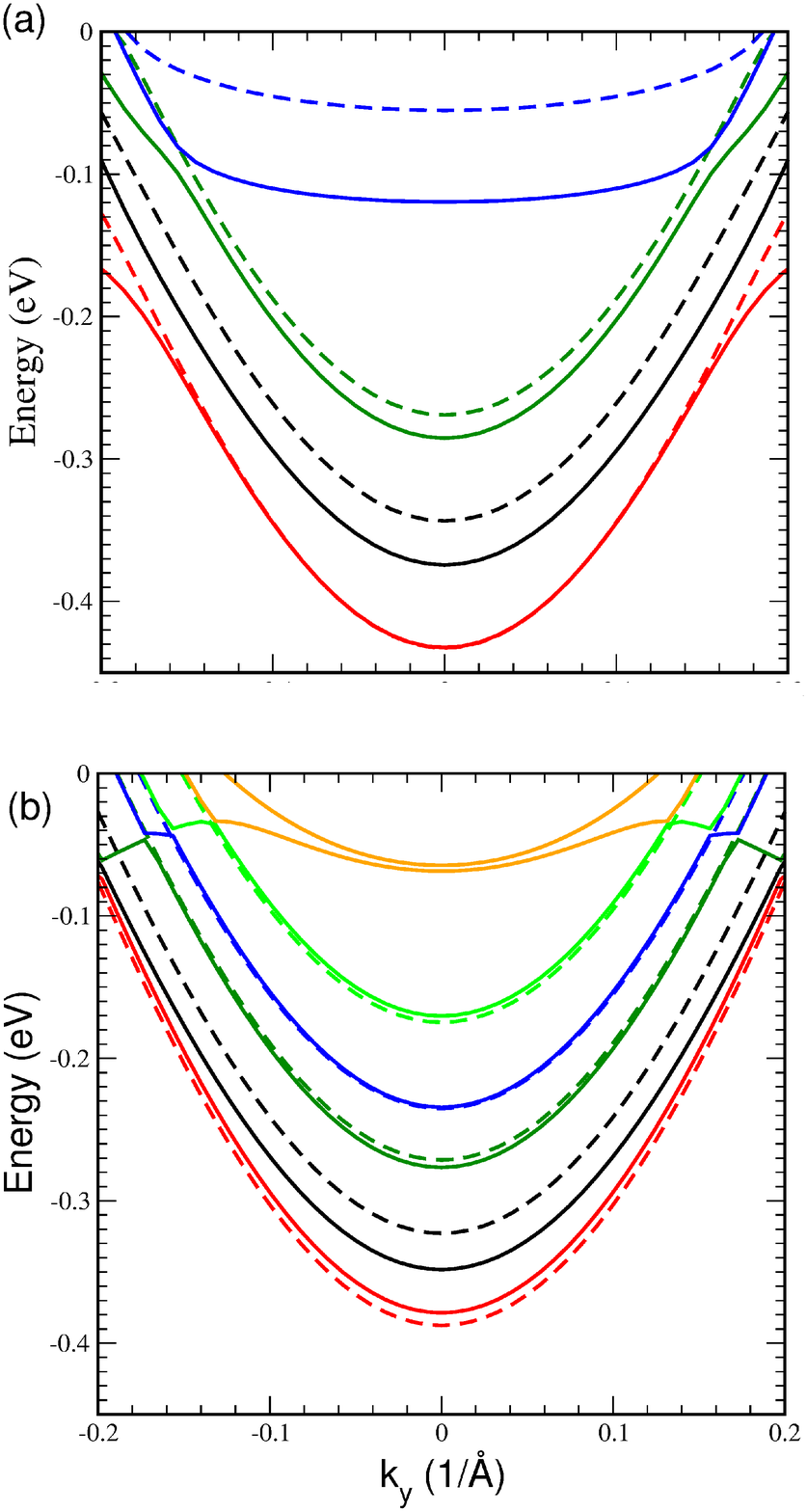}
\end{center}
\caption{(color-online) Comparison of the near zone-center conduction bands of 
  oxygen-deficient (a) 4 layer STO and (b) 6 layer STO with $n_{v} = 1/4$.
  Up-spin bands are denoted by solid lines and down-spin bands by corresponding dashed
 lines. The spin-split is maximum for the bands near the Ov surface and reduces as we go deeper 
 into the bulk.} 
\label{fig:S_4layer_6layer_gammabands}
\end{figure}

To argue that a four STO-layer slab is enough to capture the semi-infinite-ness
 of the
experimental samples, we show in Fig.~\ref{fig:S_4layer_6layer_gammabands}
the near zone-center bands of a four STO-layer and a six STO-layer slabs for $n_{v} = 1/4$ . The six STO-layer 
slab has more  subbands of  \textit{d}$_{xy}$ character, but most of these
are away from the depth-range of SARPES (as demonstrated in Fig.~\ref{fig:S_y_lm_5layer} for five STO-layer slabs), 
and the band structure near the Fermi level is the same
between the four and six STO-layer slabs. 
We therefore believe that the results obtained from four STO-layer 
thick slabs may be valid 
for the semi-infinite system.

\section{Supplementary Note 3: Orbital- and Site-Projection of Bands}
\begin{figure}[htp]
\vskip 0.3 in
\begin{center}
\includegraphics[width=\figwidth]{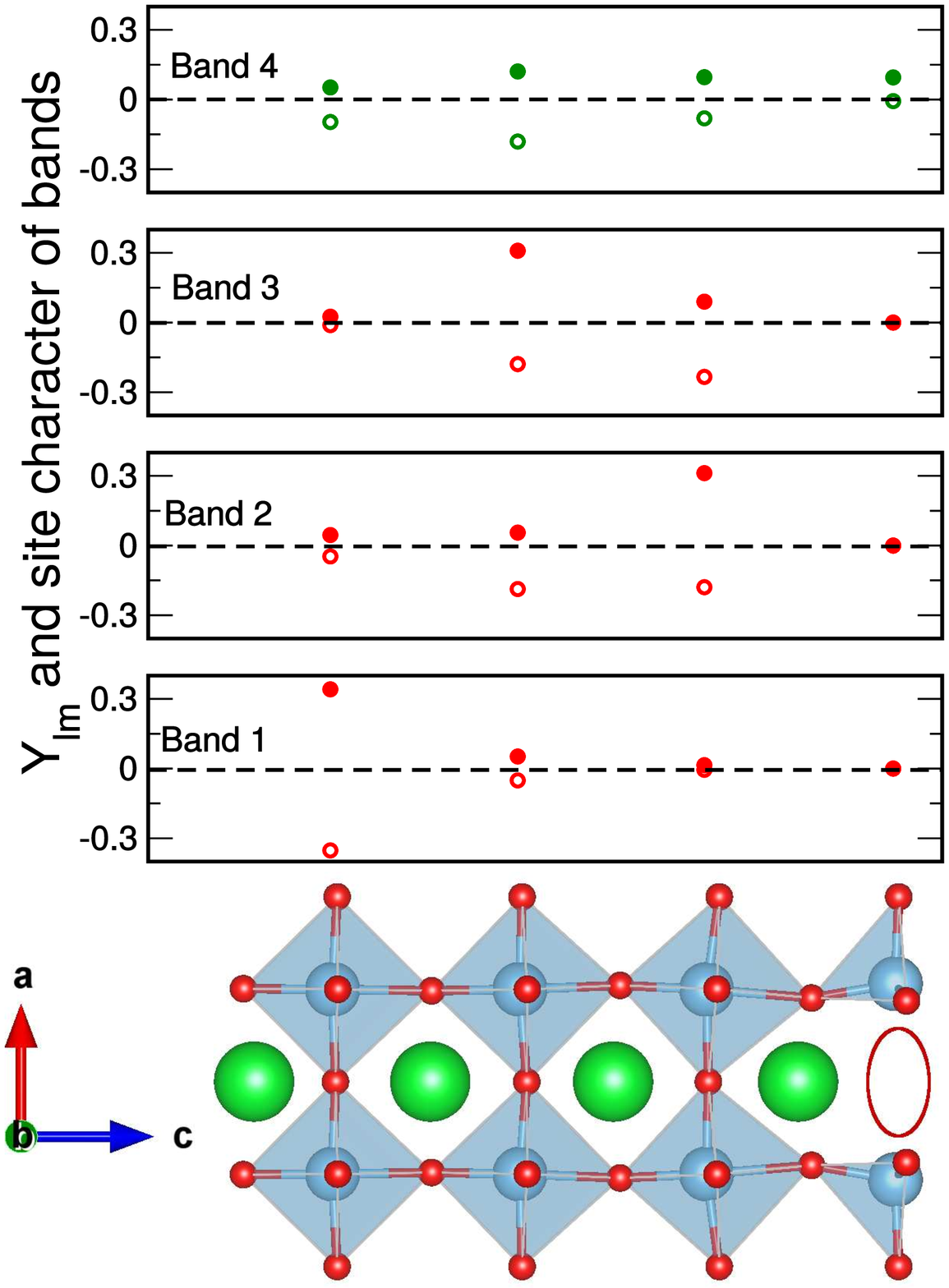}
\end{center}
\caption{(color-online) Orbital and site-projection of the 
  lowest 4 spin-pairs of bands of a 4 layer STO slab $(n_{v} = 1/4)$
   around the $\Gamma$ point and
  along $k_y$, with band 1 being the deepest pair. The site
  locations are shown by the crystal unit cell aligned along 
  the horizontal axis. On the graph, red is \textit{d}$_{xy}$ and green is \textit{d}$_{xz}$.
   Up-spin is above the axis
  and is shown in filled circles whereas down-spin is below the axis
  and is shown in empty circles. The position of the missing oxygen on the surface
  is shown with a red circle. Each \textit{d}$_{xy}$ can be
  seen to be  localized more on a distinct TiO$_2$ plane. The
  deepest pair of bands is Ti \textit{d}$_{xy}$ near the SrO
  surface and the next deepest pair is Ti \textit{d}$_{xy}$
  next to the TiO$_2$ surface. The \textit{d}$_{xz}$ band
  can be seen to be more de-localized than the \textit{d}$_{xy}$.}
\label{fig:S_y_lm_4layer}
\end{figure}

\begin{figure}[htp]
\vskip 0.3 in
\begin{center}
\includegraphics[width=\figwidth]{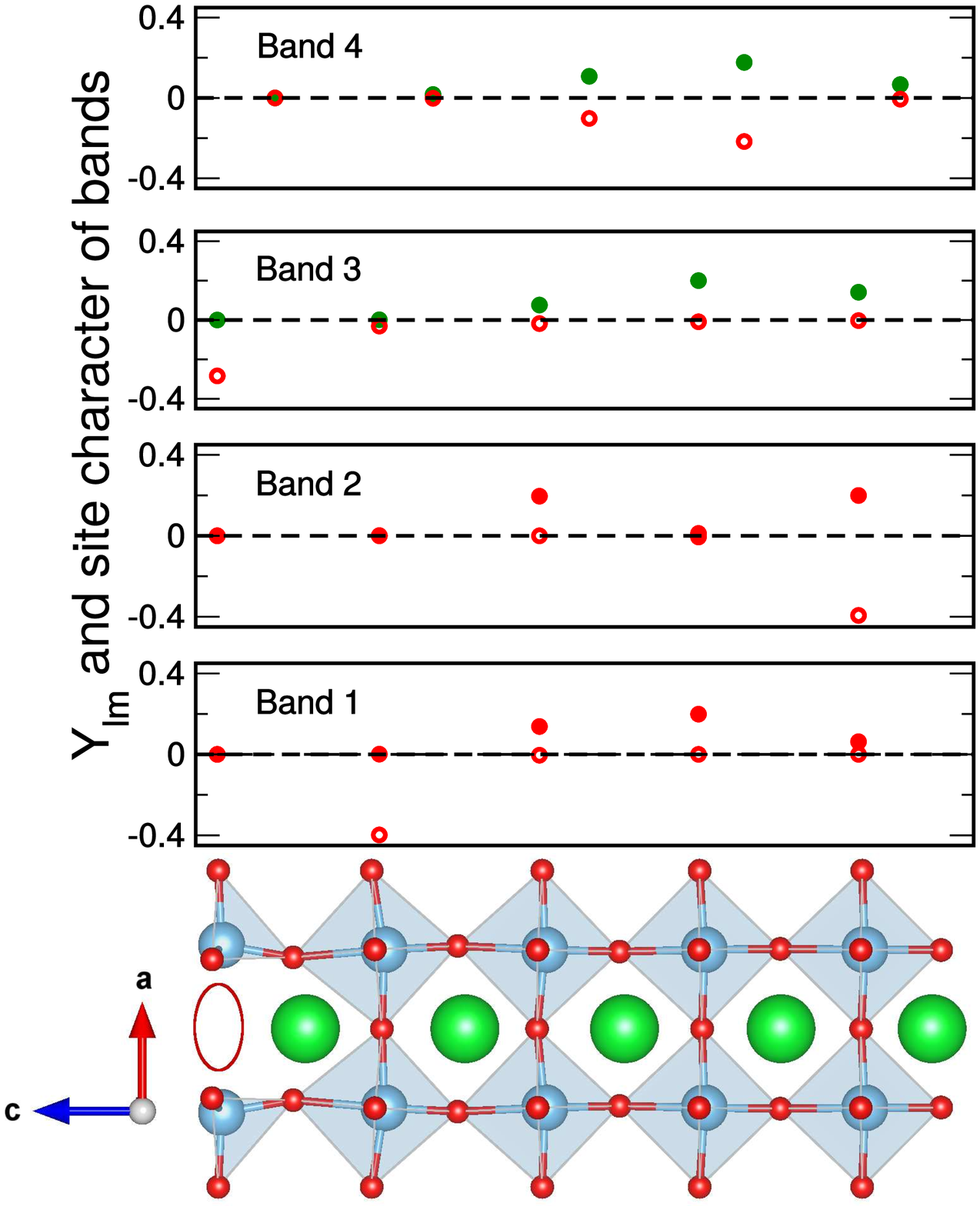}
\end{center}
\caption{(color-online) Orbital and site-projection of the 
  lowest 4 spin-pairs of bands of a 5 layer STO slab
  $(n_{v} = 1/4)$ around the $\Gamma$ point and
  along $k_y$, with band 1 being the deepest pair. A GGA+U
  scheme has been used. The site
  locations are shown by the crystal unit cell aligned along 
  the horizontal axis. On the graph, red is \textit{d}$_{xy}$ and 
  green is \textit{d}$_{xz}$. The position of a missing oxygen on the
  surface is shown with a red circle. The Coulomb term causes the spins in a pair to
  move away from each other in real space, but the quantization
  effect of the confining potential is maintained.    
   Up-spin is above the axis
  and is shown in filled circles whereas down-spin is below the axis
  and is shown in empty circles.
}
\label{fig:S_y_lm_5layer}
\end{figure}

In Fig.~\ref{fig:S_y_lm_4layer} the orbital and site-projection of the 
lowest four spin-pairs of bands around the $\Gamma$ point and
along $k_y$ for the four STO-layer slab with $n_{v} = 1/4$ 
is illustrated with band 1 being the deepest pair. The colors 
signify the orbital character of the band (red = $d_{xy}$, green = $d_{xz}$)
and the unit cell along the horizontal axis is a reference to the Ti site-location
of these states inside the unit cell. 
A Coulomb U in the form of GGA+U repels the spins from one
 another in real space and causes the spin-pairs
to spread, as a result of which the up-spin and the down-spin bands
are nested on neighboring Ti atoms, and the spin-gap is widened.
We demonstrate this in Fig.~\ref{fig:S_y_lm_5layer} which  
illustrates the near-$\Gamma$
orbital and site projection of the deepest 
corresponding occupied bands 
around the $\Gamma$-point for a five STO-layer slab $(n_{v} = 1/4)$
 slab  (using $U=4$ eV and $J=1$ eV).

\section{Supplementary Note 4: Spin-orbit coupling}
\begin{figure}[htp]
\vskip 0.1 in
\begin{center}
\includegraphics[width=\figwidth]{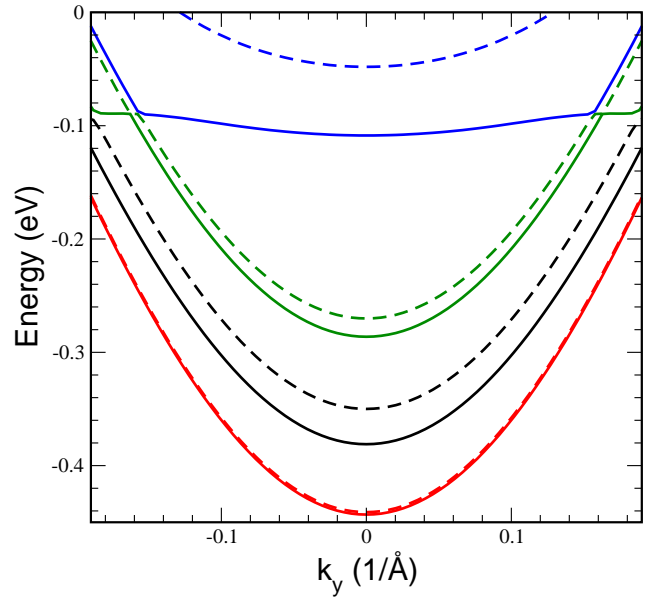}
\end{center}
\caption{(color-online) The near-$\Gamma$ band structure of the
  oxygen-deficient 4 layer STO $(n_{v} = 1/4)$ with spin-orbit coupling included.
  Up-spins are denoted with  solid lines 
  and down-spins denoted with dashed lines. The change in the \textit{d}$_{xy}$
  bands around close to the $\Gamma$ point is negligible. 
  The biggest effect of SOC can be seen in lifting the degeneracy of the
band-crossing at the corners.}
\label{fig:S_SOC}
\end{figure}

In Fig.~\ref{fig:S_SOC} we give the results obtained by switching 
on the spin-orbit coupling (SOC) in a four STO-layer slab with $n_v=1/4$. As can be seen, 
 we get a slightly modified near-$\Gamma$ band structure as compared to Fig.~\ref{fig:S_4layer_6layer_gammabands} where the
SOC was neglected.
 The effect of SOC is found to be maximum where the bands cross-each 
other, as expected.  It preserves the net magnetic  moment at the surface and the spin-polarization of the 
 \textit{d}$_{xy}$ bands around the $\Gamma$ point.

\section{Supplementary Note 5: Density of States of the T\lowercase{i} atom one layer below the surface}
\begin{figure}[htp]
\vskip 0.3 in
\begin{center}
\includegraphics[width=\figwidth]{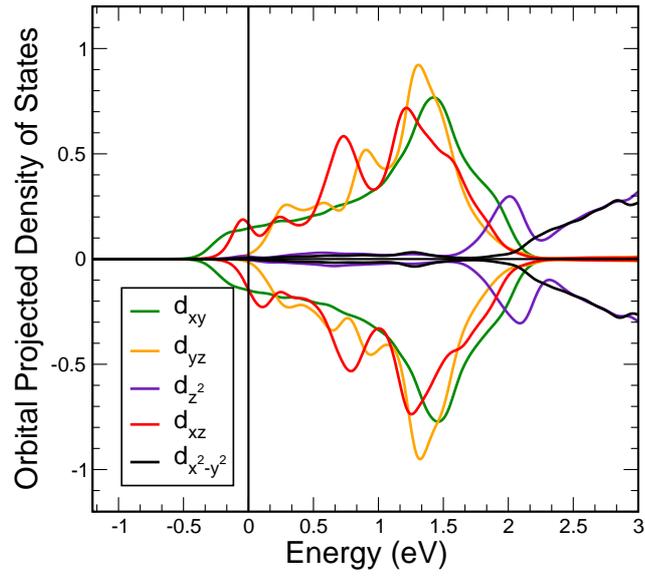}
\end{center}
\caption{(color-online) DOS of the 3d states of a Ti atom one layer
  below the oxygen-deficient TiO$_2$ surface $(n_{v} = 1/6)$. The occupied states
  around the zone-center are \textit{t}$_{2g}$ in nature. 
  At this GGA-level, the \textit{d}$_{xy}$
  states carry negligible magnetic moments, 
and most of the spin-polarization is in the
 \textit{d}$_{xz}$ state.}
\label{fig:S_Ti_3_dos}
\end{figure}

In contrast to the surface Ti atoms whose occupied states are
a mix of \textit{t}$_{2g}$ and \textit{e}$_g$, the 
projected density of states in the Ti atom one layer below the surface 
(Fig.~\ref{fig:S_Ti_3_dos}) with $n_{v} = 1/6$ shows far
less polarization, with its occupied states being predominantly
\textit{t}$_{2g}$ in character and located near the zone-center.
Here the magnetic moment is an order of magnitude less than that on the surface,
and most of the polarization is in the \textit{d}$_{xz}$ state.

\section{Supplementary Note 6: Stability of Oxygen Vacancies}
The possible range 
of Oxygen chemical potential $\mu_{0}$ can be determined by first noting that
the value is bounded from above by the chemical potential of the
 triplet O$_2$ molecule,
\begin{eqnarray}
\mu_{O} < \mu_{O_2}/2 = -4.917 \mathrm{eV}. \hskip 1.3 in \mathrm{(S.1)}\nonumber 
\end{eqnarray}
Furthermore, it is constrained by the relation
\begin{eqnarray}
\mu_{Sr} + \mu_{Ti} + 3\mu_{O} = \mu_{SrTiO_{3}}^{bulk} = -40.125 \mathrm{eV},
\hskip 0.2 in \mathrm{(S.2)}\nonumber 
\end{eqnarray}
 where the chemical potentials of Sr and Ti must themselves satisfy inequality constrains:
\begin{eqnarray}
\mu_{Sr} < \mu_{Sr}^{bulk} = -1.6840 \mathrm{eV}, \hskip 1.2 in \mathrm{(S.3)}\nonumber \\ 
\mu_{Ti} < \mu_{Ti}^{bulk} = -7.8981 \mathrm{eV}. \hskip 1.2 in \mathrm{(S.4)}\nonumber
\end{eqnarray}
Therefore the allowable energy of the oxygen chemical potential is 
\begin{eqnarray}
-10.181 \mathrm{eV} < \mu_{O} < -4.917 \mathrm{eV}. \hskip 0.9 in 
\mathrm{(S.5)}\nonumber
\end{eqnarray}
The free-energy per unit cell plotted in Fig.~2 for each oxygen deficient structure is taken to be
\begin{eqnarray}
F = \Bigl [ F_{V} - F_{0} + \mu_0 \Bigr ] /N_u, \hskip 1.2 in \mathrm{(S.6)}\nonumber
\end{eqnarray}
where $F_{V}$ is the free-energy of the oxygen deficient structure, $F_0$ is the free-energy of the no-vacancy structure
and $N_u$ is the number of $1 \times 1$ unit cells needed to make the supercell used in our DFT.

All three structures are stable at only very low oxygen chemical potential in equilibrium.
Therefore, we believe that non-equilibrium effects like intense UV irradiation is necessary to
create oxygen vacancies with the density required for the spin-split. 

\begin{table}[H]
\begin{center}
\begin{tabular}{ |c|c|c|c|c|c|c| } 
 \hline
 \textbf{Ov type($n_{v}$)} & stripe(1/4), & stripe(1/6), Ti(1) & stripe(1/6),Ti(2) & stripe(1/8) & dimer & $2\times1$ chain \\
  \hline 
  \textbf{surface} & 0.391 & 0.118 & 0.228 & 0.000 & 0.368 & 0.000\\ 
  \textbf{1 layer deep} & 0.046 & 0.030 & 0.069 & 0.000 & 0.033 &0.000 \\ 
 \hline
\end{tabular}
\caption{
       Magnetic moments of surface Ti atoms and of those one layer
       below the surface, in units of $\mu$B. For $n_{v} = 1/6$, Ti(1) refers to the atom closest
       to the O vacancy, and Ti(2) refers to the atom farther away. 
}
\label{table:moments}
\end{center}
\end{table}

\newpage
  
\begin{figure}[htp]
\vskip 0.1 in
\begin{center}
\includegraphics[width=\figwidth]{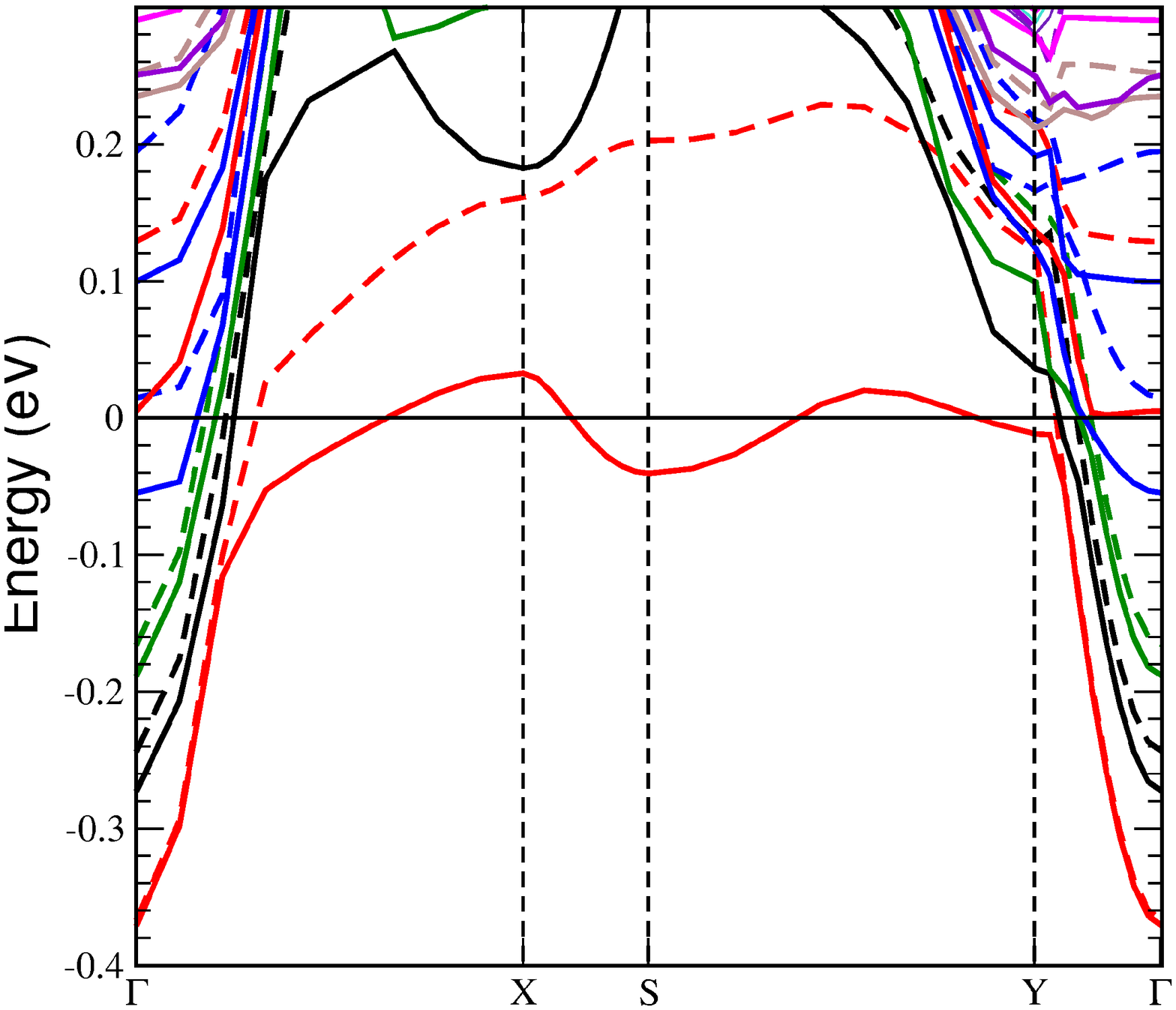}
\end{center}
\caption{(color-online) The planar BZ band structure of the
  oxygen-deficient four STO-layer slab with $n_{v} = 1/6$.
  Up-spins are denoted with  solid lines 
  and down-spins denoted with dashed lines. The lowest occupied pair
  of spin bands is degenerate near the $\Gamma$ point and is 
  contributed by the Ti atoms away from the oxygen deficient TiO$_2$ surface. 
The second lowest
  spin-split payer comes from the Ti atoms one layer below the oxygen deficient
  TiO$_2$ surface, which are the bands of interest. The majority of the magnetic moment
  comes from the BZ corner and edge where the states are localized on the doped TiO$_2$ surface
  as a mixture of $t_{2g}$ and $e_g$ with a considerable spin-split.}
\label{fig:S_4layer_fullbands_1_6}
\end{figure}
\bibliography{sohambib}